\documentclass[11pt,journal,compsoc]{IEEEtran}
\usepackage{graphicx}
\usepackage{amsmath}

\begin{document}
%
\title{Diagnosing Distributed Systems through Log Data Analysis}

\author{\IEEEauthorblockN{ \textsuperscript{1.} K R Chowdhary, Professor  \textsuperscript{2.} Rajendra Purohit, Associate Professor}\\
\IEEEauthorblockA{Dept. of Computer Science  and Engineering\\
Jodhpur Institute of Engineeering and Technology, Jodhpur \\
Email: 1, 2. kr.chowdhary, rajendra.purohit@jietjodhpur.ac.in}}

\maketitle

\begin{abstract}
The log-based analysis and trouble-shooting has remained prevalent and commonly used approach for centralized and time-haring systems. However, for parallel and distributed systems where happen-before relations are not directly available between the events, it become a challenge to fully depend on log-based analysis in such instances. This article attempts to provide solutions using log-based performance analysis of centralized system, and demonstrates the results and their effectiveness, as well presents the challenges and proposes solutions for performance analysis in distributed and parallel systems.

\end{abstract}

\IEEEpeerreviewmaketitle

\section{Introduction}

Failure analysis as well as the health monitoring of a computing facility is as important as its actual working. Apart from failure analysis, often there is needs to investigate the systems for other factors, e.g., to find out whether there is malware in the system, or when some unauthorized user or program accessed the system and logged-out; we are keen to investigate the time of such login, changes it has caused in the files of the system, and programs it has run. Similarly, we as administrator are keen to know about the failed attempts made by unauthorized program/user for  accessing any system from remote locations. Other reasons of investigation may be: we are keen to know, e.g., what events have caused the hard-disk to crash or have caused the CPU to go into deadlock? Yet other motivations for log-based analysis may be: to do the performance study -- are processors over or under-loaded, or does IO (Input-output) have sufficient band-width, or if there is too much of swapping between RAM and storage, is configured virtual memory sufficient, or many similar factors.

Diagnosis of centralized system is often needed to trouble-shoot it to investigate various types of failures, which are related to IO, process, memory, cache, CPU, etc. Because the propagation delays are more or less fixed in the centralized system, and can be estimated accurately, the sequence of events recorded in log accurately map to the order of occurrence of events; an event may be a cause or an effect. Accordingly, the facility of systems logs are helpful tool for performance analysis as well as trouble shooting the failures in the centralized parallel systems. However, due to propagation delays and varying nature of these delays in distributed systems (DS), it becomes a challenge to analyze and investigate the faults in such systems.

This paper presents the potential solutions for centralized systems that are based on Unix/Linux, through log analysis, to investigate the health and performance of systems. However, typical DS logs do not comprise enough information to simulate {\em happens-before} relation between events, hence they cannot be interpreted in time frame reference. This paper explores various techniques, and proposes methods to analyses the events in a distributed system, supported with analysis for typical scenarios. 

The organization of this paper is as follows.The section II presents log-based analysis of system using {\em system-log}, and explains how these logs, which are in large size, can be searched at speed; next it presents the challenges in multi-threaded system. The Section III discusses how to carry out the performance analysis and performance predictions, section IV presents approaches and methods to be used for security related issues, and what and how the logs available for these. The section V suggests the tools as  well as the limitations of distributed systems analysis and how to make use of distributed time-stamps, and section VI presents the conclusions of this article.

\section{Log based analysis of Systems}

The logs stored in a computer's logfile provide the glimpses about the running system. The logs are short messages that are collected in system specific logs files, however, their format and contents vary from system to system. Some of the examples of the log are: trouble in communicating with printer, details of pages demanded by client from the web-server that are stored in web-server, a messages about hard-disk access, etc. These logs have important applications: e.g., web server-log about the requests made to server may give information about traffic patters on the server, printer-log may indicate the frequency of requests to printer, and hard-disk read/write log may give information about speed of hard-disk, its response, load on it, latency of the hard-disk, etc~\cite{adam12}.

Operating system log data can provide healthy diagnostic information about computer and networks, and Linux is no exception. All the events from kernel events to user applications and commands are logged by Linux, allowing us to see almost any action performed on computer system. Logs provide a visual history of series of events that has been happening in the heart of a Linux operating system. So, if anything goes wrong, logs give a detailed list of events in order to help user and explains what Linux logs are, at what place user can find them, and how to interpret them. Systematic logging placement is desired to measure performance overhead and trivial logs that include important runtime information. Deciphering is easy for Linux Log file as it is stored in the directory ``/var/log" and its subdirectories, in text form; this log consists of kernel, events, package manager, etc. In Linux, /var/log directory is a special directory, in which logs from services, the OS itself, and applications running logs on the system are included~\cite{Johan02}.

\subsection{Linux Log}
 
In today's log system, in spite of many advancements, the \texttt{printf} is used to log the messages to console or local disk. The \texttt{printf} is used along with some manual inspection and {\em regular expression} commands to locate the specific command patterns in the log file. The simplest of these is \texttt{grep} filter command for specific messages. The system administrator has to anticipate the possible reason, say, the system crashed due to network failure. The admin may check it by,\medskip

\noindent
\texttt{\$ cat syslog | grep "connection dropped"}\\

\noindent
that is, by searching into the system log, which will provide the listing of all such lines. 

\begin{figure}
\centering
\includegraphics[scale=0.69]{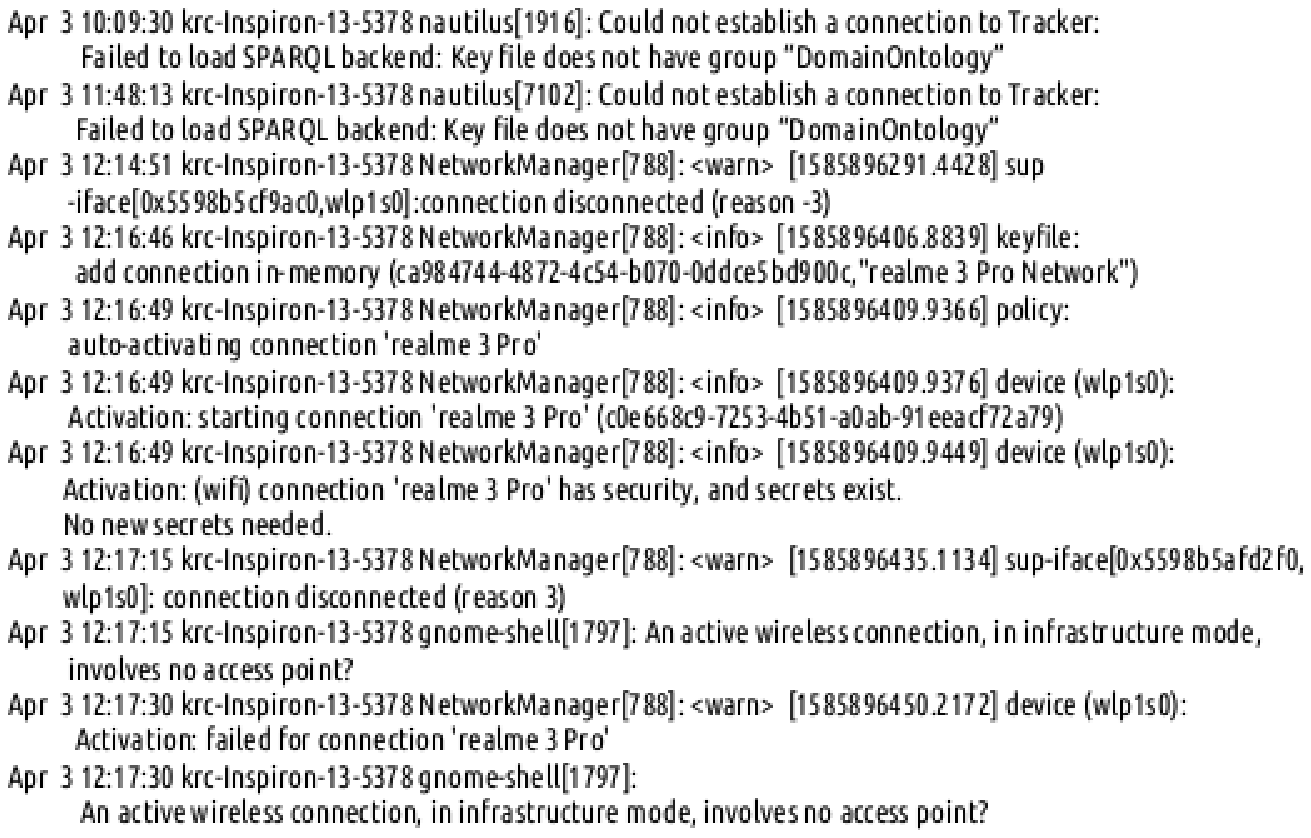}
\caption{System log for keyword `connection' from Linux}
\label{fig:syslog}
\end{figure}

To find out the information about various type of connections the system has made, we simply use keyword 'connection' in the above mentioned command,\medskip

\noindent
\texttt{\$ cat syslog | grep "connection"}\\

\noindent
which generates the output from syslog, part of that is shown in Fig.~\ref{fig:syslog}:\\

Analysis of this log can be helpful in many ways: In the beginning it says that connection could not be established to Tracker At 10.09:30 (April 3); further as logical explanation it says, ``Failed to load SPARQL backend: key file does not have ``DomainOntology". Later, at 11:48:13, probably the connection was lost, and again there is log entry indicating the same error. At 12:16:46, the log says about adding connection, to `realme 3 Pro' network, which is due to hotspot of a smartphone. From this fraction of the log we are able to analyze the health of wireless connection provided to the Linux system, provided through smartphone hotspot. Every event is specified with details, like: time of event, name (id) of the system, kind of the fault, and consequence (effect) of that on the system/software. 

When we think of a particular issue in the system, there is need of estimating a suitable keyword(s) which are likely to have mapping with the specific trouble/fault in the system. Hence there should be an entry in {\em syslog} with that keyword embedded. And, those lines of the log can be dumped on console using, e.g., {\it grep} command of linux to investigate the true cause of the fault in the system.

However, when we consider another situation, e.g., ``web service suddenly becomes slow," the console operator is not likely to see this error, stating that ``latency in service request increased by $10\%$ due to bug $X$ triggered!" Thus computer operators are likely to search the keyword ``failure" or ``error". Such keywords are used inaccurately, the developers who write the code of logs management, have no detailed knowledge about the keywords that will ultimately be used.

\subsection{Multithreaded Systems}

The logging process is with reference to some internal synchronization. The logging process in multithreaded systems is complex as the debugging of such systems is challenge due to continuous change in interleaving patterns of threads. The key observation in such scenarios is non-deterministic program behaviors at certain execution points, e.g., interrupts by clock and by input/outputs. Possible solution is, to log all these nondeterminic execution points, so that it becomes possible to accurately replay the entire program. The replay process of program is powerful because using this technique one can observe every thing in the program by modifying the instrumentation prior to the relay. However, for the programs of concurrent nature or those programs where deterministic execution depends on large amount of data, such methods are not practicable.

Another challenge is that log size can be excessively large in multithreaded systems, for example, one may want to log every ``acquire-release" sequences on a clock object in order to debug the log contention. This can be taken care of by heterogeneous logs. But such logs will require further analysis. In addition, there is definite cost in storing, sorting, and indexing large amount of messages, to be retrieved or searched later, but many of these logs may never be used.

The other approach to maintain the logs is to store only the aggregated or summary information and not every message/event. These logs may be helpful to discover more detailed information about the system's performance using machine-learning based analysis, e.g., through PCA (Principle Component Analysis) and SVM (Support Vector Machines). These approaches are important in networked or distributed systems, where collecting even a single value from each component may result to large performance cost. The machine learning techniques may be used to discover useful log messages in the log files, these techniques may be based on the criteria of anomaly detection. For machine learning techniques to be applied, the input data is treated as feature vectors. However, it is a nontrivial to convert free-text log messages to meaningful features~\cite{chowdhary20m}.

There are challenges in statistical anomaly detection due to the fact that there are no ways available as evidence to find out in case of some messages, as whether they are {\em causes} or {\em effects} (symptoms). For example,``read failure" of a disk sector is ``cause" in the error message ``load failure", while it is a symptom as per log entry, as there is evidence of hard-disk read failure.

The static analysis of program can help in discovering the root cause of the messages in the log. This may be possible by analyzing the path in the program that could lead to the message. The approach can also be helpful in improving quality of log by finding the divergence points; it may be possible that the program had entered the error path through these points.

\section{Performance Analysis}

The log analysis can be of much help in optimizing the system performance. To improve the system performance, it is important to understand how the system resources are used. Some logs can track how individual resource is used, and can provide the time series. Statistics of resource usage is measured in the  form of cumulative usage per unit time, e.g., $x$ bits transmitted in last $20$ secs., or $100$ instructions executed in last $5$ micro-secs., etc. The bandwidth parameter can be used to represent the disk or network performance, or number of page-swaps, to represent effectiveness of the memory. In similar ways, CPU scheduling can be used as a characteristic property for load balancing. 

\begin{figure}[h]
\centering
\includegraphics[scale=0.8]{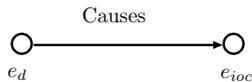}
\caption{Cause effect relation between two events}
\label{fig:infr}
\end{figure}

The performance problems occur due to interaction among different system components; to reveal these interactions we need to synthesize the textual information in the logs generated by multiple resources. However, the components in {\em distributed systems} may not agree on a uniform format of these log messages, therefore the log format is invariably heterogeneous. Due to varying propagation delays, and due to heterogeneous format of logs, determining the exact order of events from multiple components is almost an  impossible task. In addition, an event of one component, e.g., event of flushing of data by disk ($e_d$), may cause serious problem to other resource, e.g., contention of I/O resource (event $e_{ioc}$). Because, when a component causing the problem is not likely to log the event, then it may be a challenge to find out the root cause (of I/O contention). An approach to solve such problem is, look into the interrelated behavior of components, and compute the influence to infer the relationship between components or group of components. The Fig.~\ref{fig:infr} shows form of a simple cause effect relation, of these two events.

The relationship between these two events can be expressed using Bayes theorem to derive the inference, and expressed in the form of equation, i.e., what is probability of occurrence of $e_d$ given that $e_{ioc}$ has been observed. This relation is expressed by (\ref{eq:bayes1}),

\begin{equation}\label{eq:bayes1}
P(e_d / e_{ioc}) = \frac{P(e_{ioc}/e_d) \times P(e_d)}{P(e_{ioc})}. 
\end{equation}

The cause effect relationship of Fig~\ref{fig:infr} can be extended to a chain of cause-effect relations, and can be represented using a directed graph $G = (V, E)$, with set of nodes (vertices) $V$ and set of edges $E$. Having representation as a directed graph, using equation~(\ref{eq:bayes1}) it is possible to compute the causes which resulted to chain of consequences and has resulted to specific symptoms~\cite{chowdhary11}.

One of the challenges in log analysis is risk of influencing the measurements due to the act of measuring. Excessive degree of logging the events may consume resources, e.g., some times eating the entire hard-disk space in a short time, which may also make the measurements more complicated. The more we measure, there will be lesser scope for deriving and understanding the system's performance characteristics. To reduce the impact on performance due to excessive logging of events, an approach can be sampling of events. However, this will make provisions for missing certain rare and important events.

One tool for distributed log analysis is {\em DTrace}\footnote{The DTraceToolkit is a collection of useful documented DTrace scripts. There are over $200$ scripts in the DTraceToolkit, and each has a man page and an file of example output.}, that requires statistically instrumented log sites. Other, toolkit, called, {\em Fay}~\cite{erling12}, provided a platform for collecting, processing, and analysis of program execution traces. This tool permits the users to specify the list of events, which need to be measured. These specifications are in the form of certain query formats of declarative language. The tool then inserts dynamic instrumentation in the running system, which aggregates the measurements, and provide query specific analysis mechanisms.

\subsection*{Performance predictions}

The log data can be used for provisions for the future, i.e., about providing of resources based on the past statistics, e.g., planning the capacity of the system, management of workload, scheduling executions, and to configure the system optimization. Prediction models can be designed for I/O subsystems, hard-disk arrays, databases systems, and Web servers, which may be based on probabilistic models, like, given in (\ref{eq:bayes1}) above.

The predictions require extraction of feature vectors from events logs -- a nontrivial process, but critical, which affects the effectiveness of the predictive model. The predictive models often provide a range of values that indicate the confidence interval, rather than a single value.

\section{Security Analysis}

The logs maintained by the system are also useful for security related usages, e.g., for investigating the security breaches, like intrusion detection or misbehavior of the system itself. This is carried out in the form of postmortem  inspection based on the log contents. However, the effectiveness of logs depend on system and threat model used. Various security related logs are: logs of system logins sessions, of firewalls, log of resource utilization, e.g., system calls, network flows, etc.  

\begin{figure}
\centering
\includegraphics[scale=0.6]{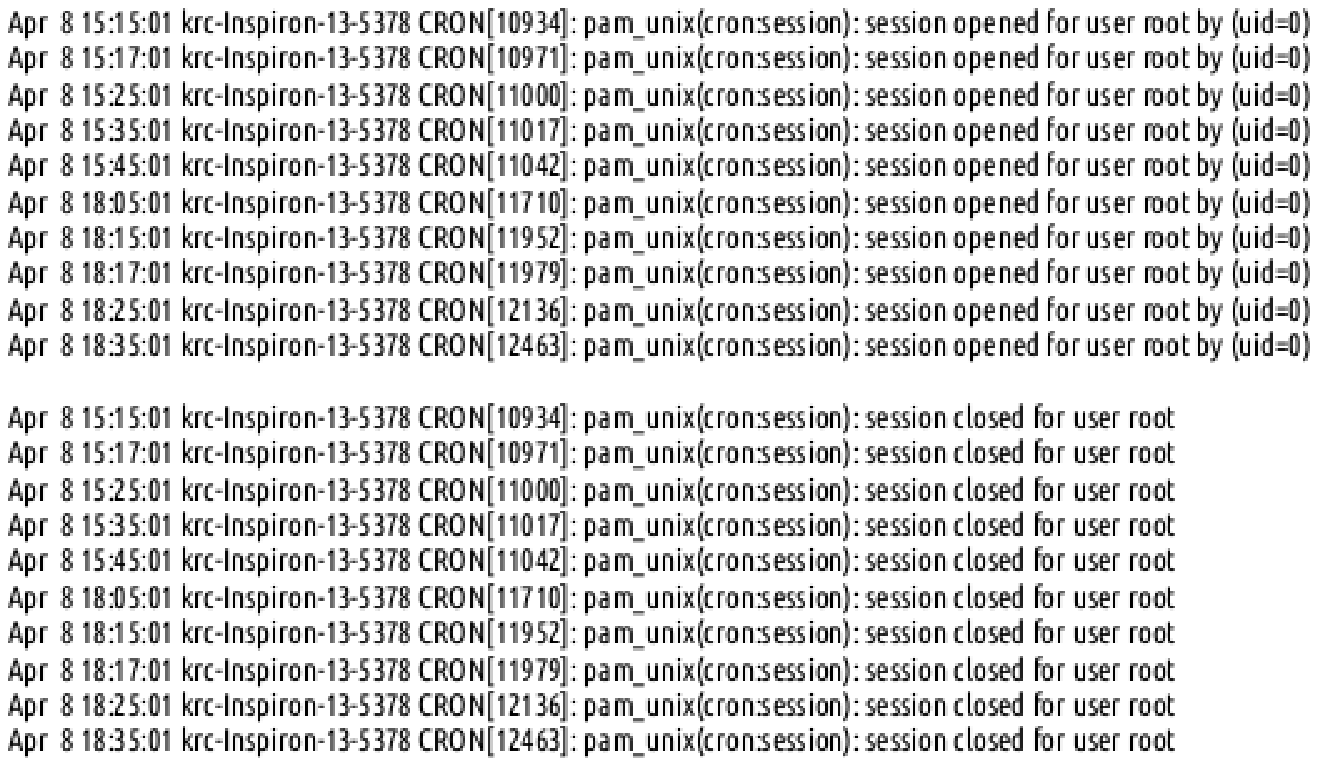}
\caption{Log of a session opened and then closed by intruder as root}
\label{fig:logscoroot}
\end{figure}

Where a user logs into machine remotely via SSH (secure shell), the host machine generates log entries corresponding to login event. To find out the details about unauthorized access to system resource, the commands listed  below lets us know about the session opened by root, and after closed by root, with in between activities carried out by this user. The corresponding logs results are shown in Fig.~\ref{fig:logscoroot}. The time-stamps in these lets us know, if the intruder acted as root or it is a genuine root user.\medskip

\noindent
\texttt{\$ sudo cat /var/log/auth.log | grep "session opened for user root"
}\medskip

\noindent
\texttt{\$ sudo cat /var/log/auth.log | grep "session closed for user root"}\medskip

We briefly introduce the logs of Unix/Linux before we proceed to explore their applications for system security. These logs called {\em system logs}; they are primarily concerned with functioning of the system, and are not necessarily the additional applications added by the developers or users. The examples of system logs are: system daemons, authorization mechanisms, and system messages, which are all maintained in the file {\em /home/var/syslog}.

\subsection{Authorization Log}

\begin{figure}
\centering
\includegraphics[scale=0.62]{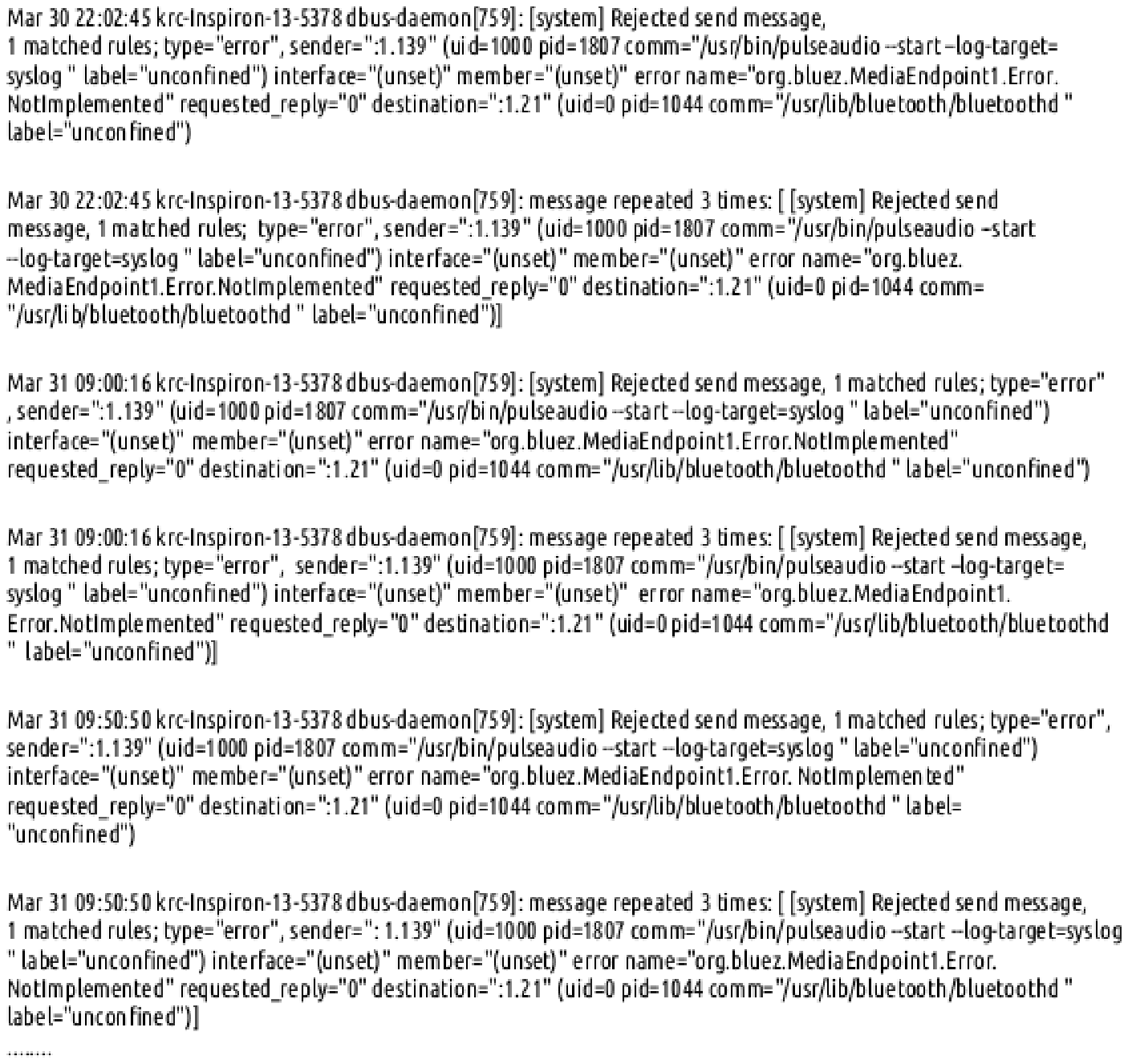}
\caption{Header of authlog having "error" keyword}
\label{fig:authlog}
\end{figure}

It tracks the usages of {\em authorization systems} of Unix, i.e., tracks the mechanisms which authorize the system users by prompting them for user passwords. The examples of such systems are: Pluggable Authentication Module (PAM), \texttt{sudo} (super-user do) command, and remote login to \texttt{sshd}. The authentication log (/var/log/auth.log) is useful for learning about user logins and to know about sudo command's usages history. For example, to find out the traces of \texttt{sshd} logins, we can filter the contents of this log and dump the header of the entries carrying keyword ``error" (output shown in Fig.~\ref{fig:authlog}) using the \texttt{grep} filter, as follows:\medskip

\noindent
\texttt{\$ grep "error" /var/log/auth.log | head}\\

\subsection{Daemon Log}

The {\em Daemon} program runs in the background mode, often with no user intervention, and performs some operations that are essential for the proper functioning of the system. The daemon log (/var/log/daemon.log) contain the run time information of the system, and about some of the applications, like  Gnome Display Manager, bluetooth HCI, or MYSQL daemon. This log is helpful in trouble-shooting the problems related to the respective daemons.

\subsection{Debug Log}

The debug log (\texttt{/var/log/debug}) provides messages from Linux system and applications running under this, which are helpful for debugging the system configuration as well as for debugging the applications. 

\subsection{Kernel Log}

The log messages about the health of all the tasks performed by the Kernel are logged into /var/log/kern.log, which may prove useful for analysis and trouble-shooting the kernel of the system.

\subsection{System Log}

The system log (/var/log/syslog) records the largest amount of information about the entire system, which includes processes, web access, resource utilization, core/package temperature and speed, etc.   

\subsection{Application Logs}

Two examples of application logs are due to Apache2 web-server (/var/log/apache2) and Sambha Server (/var/log/samba). Corresponding to these, there are two files in default sub-directories: 1. /var/log /apache2/access.log, which records every file loaded and every page served by the web server, and 2. /var/log/apache2/error.log records all the error conditions reported by HTTP server. A default log (/var/log/cups/error\_log)  is used by Common Unix Printing System (CUPS) to log the information and messages. Whenever there is a printing issue in Unix, this log is proper place to start for troubleshooting.
 
In addition, the SMB (server message block) protocol server is used for sharing files between Unix and other computers which support SMB protocol. The Samba maintains three different logs in /var/log/samba. The other logs are shown in Table~\ref{tab:logs}.

\begin{tiny}
\begin{table}
\caption{Logs and their functions in Linux}
\label{tab:logs}
\begin{tabular}{|l|l|l|}
\hline
Name of Log & Path of log file & Functions \\
\hline               
Login Records Log & \texttt{/var/log/wtmp} &  Currently logged\\
                  &                        & in users\\
\hline
Login Failure Log & \texttt{/var/log/faillog} & Login failures \\
\hline                  
Last Logic Log & \texttt{/var/log/lastlog}  &  Last logins\\
\hline 
\end{tabular}
\end{table} 
\end{tiny}

\subsection{Log exploration Tools}

There are some commonly used commands to work with log files in Unix/Ubuntu.

To make the log directory as current directory:\\

\texttt{\$ cd /var/log}\\
 
Viewing files:\\ 

\texttt{\$ less auth.log}\\

\texttt{\$ more auth.log}\\

\texttt{\$ head -n 10 auth.log}\\

\texttt{\$ tail auth.log}\\

\texttt{\$ tail -n 10 auth.log}\\

\texttt{\$ grep "system" auth.log}\\

\texttt{\$ grep "system" auth.log | less}\\

In spite of so many logs and tools to explore them, only some of the performance related questions can be answered using the log data and search tools. There are quite likely number of failed login attempts for an intruder, before any successful login attempt, which are not available in the the system logs. Thus, power of analysis is limited to the extent of available information in these logs. The developers can add to the incomplete logs to improve the coverage, so that it is not easy for adversaries to escape from leaving the traces of their activities~\cite{adam12}.

\subsection{Log Rotation}

The /var/log comprises logs files like, auth.log, auth.log.1, auth.log.2.gz, dpkg.log, etc, with root as owner of these files. These logs are automatically renamed after a predefined time, and new log is started with the original name. With longer times, the log files are compressed using $gzip$ utility, and renamed with extension $gz$. This process is called {\em log rotation}, which retains the old logs due to their obvious importance, while reducing the disk space due to compression. Generally, the logrotate is called from cron-script by /etc/cron.daily/logrotate utility, and its finer configuration is defined in the file /etc/logrotate.conf, which can be edited by the administrator as per needs.

\section{Log analysis for Distributed Systems}

All Distributed system (DS) pose almost unique challenges for designer of these systems. These challenges prevail due to difficulty in correct identication of system events, in the frame of reference of time, incluing the identification of topologies. Consequently, the log analysis is tedious and complex task. The factors common in all distributed systems attributed to these are: highly non-uniform traffic scenarios, and non-uniform and uncertain pattern of switching-on and switching-off of intermediate stations during the communication of any two parties. The above factors necessitates the alternation of communications paths during the middle of communication process, which leads to highly dynamic and non-deterministic nature of communication in distributed systems, leading to highly varying nature of propagation delays~\cite{Ivan16}, \cite{Liu}.

\begin{figure}
\centering
\includegraphics[scale=0.7]{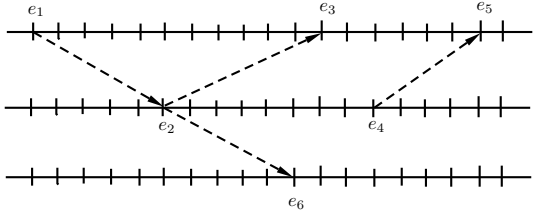}
\caption{Propagations in Distributed System}
\label{fig:pd1}
\end{figure}

Often the approach used to know about the distributed system activity is to analyze log data. Unfortunately, this is not an easy task, as typical DS logs do not comprise the information to the extent that using that one can simulate the \emph{happens-before} relation $e_i \to e_j$, where an event $e_i$ occurs before the event $e_j$. Hence, these events cannot be interpreted in reference of time frame. The Table~\ref{tab:evntdisys} shows the propagation delays between events in a typical DS whose events are graphically represented in Fig.~\ref{fig:pd1}. In this table, e.g., time of event $T_{e_2}$ is equal to time of event $T_{e_1}$ plus propagation delay (PD) from event $e_1$ to event $e_2$ ($T_{e_2} = T_{e_1} + PD_{1 \to 2}$).  

\begin{table}
\centering
\caption{Events Displays}
\label{tab:evntdisys}
\begin{tabular}{|lll|}
\hline
$T_{e_2}$ & = & $T_{e_1}+PD_{[1 \to 2]}$\\ 
$T_{e_6}$ & = & $T_{e_2}+PD_{[2 \to 6]}$\\ 
$T_{e_3}$ & = & $T_{e_2}+PD_{[2 \to 3]}$\\ 
$T_{e_5}$ & = & $T_{e_4}+PD_{[4 \to 5]}$\\
\hline
\end{tabular}
\end{table}

Simultaneous operation by multiple nodes in a distributed system allows concurrent operations, which allows parallelism in distributed system. However, the concurrence operations may create {\em deadlocks} and {\em race conditions}, which are not so easy to detect and diagnose in a distributed system. The {\em distributed algorithms} are helpful in removing potential inconsistencies; most distributed algorithms provide an assurance of some data consistencies and cache coherence. The distributed systems work on the global state of the system instead of local.  

The distributed systems are robust, and can survive the partial failures. However, it requires the developers of the distributed systems to reason through complex failure modes to achieve such fault-tolerance.

\subsection{Distributed Timestamps}

The distributed systems are event driven. Given any two events $e_i$ and $e_j$, the requirement of a distributed system is to find out the {\em happen-before} relation between them, i.e., which events occurs before what events? The Table~\ref{tab:evntdisys} shows the relative times of occurrence of five events $e_1, ..., e_5$. However, it is difficult to compute the happen before relations between the events due to varying propagation delays (PDs) caused due to varying traffic scenario in the network.
 
One approach makes use of {\em vector-clock} timestamps, where each node maintains a vector of logical clocks. For a network of $n$ nodes, a vector clock for each node is a vector of size $n$, say, $[t_1, t_2, ..., t_i, ..., t_j, ..., t_n]$. If this vector is for $i$th node, then $t_i$ is logical clock of local node, i.e., $i$th node. A message sent by a node also comprises its part, the vector clock of seding node. Each time $i$th node sends a message to some other node $j$ or receives the message, or performs some local action, it increments the vector element of its own clock. When, a node receives the message from $i$th node, the receiving node updates its own vector clock -- replaces each element of its vector clock by maximum of the local copy vs received timestamp value. And, this vector timestamp becomes the part of every logged event, node-wise in a distributed network~\cite{Ivan16}. The logical clock has no direct relation to real-time clock, except that they increase in positive directions. However, if no event takes place at a node, its vector clock does not change. The following are the tasks which are possible due to vector clock.

\begin{enumerate}
\item {\em Log analysis.} This is a light weight approach, which works with non modifiable systems. Log analysis are the approaches, that are used to understand system log, debug log, console log, similar to Unix/Linux based systems.

\item {\em Visualization.} The density and complexity together, of distributed systems have opened the scope of visualization of systems, which is transparent to developers. The developers use visualization to understand the basic communication patterns in the distributed systems, particularly the ordering of events/messages in these systems.

\item {\em Visualizing DS Executions.} The DS executions can assist developers to comprehend, debug and visualize DS. One such tool, the ShiViz visualization tool~\cite{Ivan16}, displays system executions in the form of event diagrams, that show interactions as well the ordering of events in distributed systems, that are related by {\em partial-ordering} relations. A typical ordering is shown in Fig.~\ref{fig:pd1}. This visualization is created using the events and interactions that have been logged in execution logs. It is also possible to enlarge, disintegrate, and hide parts of events. Using these visualization it becomes possible to diagnose the events for particular interaction patterns. 
\end{enumerate}

The ShiViz tool logically explores the happens-before relation between logged events. However, it does not take care of ordering of all events, as it cannot establish the happen-before relation between every event to every other event, as some events might have happened at other nodes for which it cannot determine the logical-clock time, in the absence of communication with them. This scenario can be allowed as long as the node not logged does not effect the behavior of other events.

\begin{figure}
\centering
\includegraphics[scale=0.8]{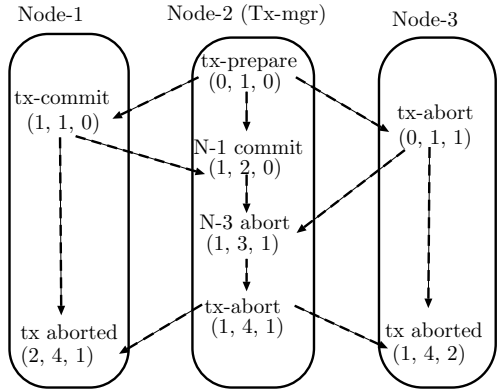}
\caption{Typical scenario events' sequence in a Distributed System}
\label{fig:sds2}
\end{figure}

An execution of {\em two-phase commit}-protocol (2PC)\footnote{The Two-phase commit protocol is an atomic commitment protocol, which makes use of distributed algorithm to coordinate processes participating in distributed atomic transactions. The end result is decison of commit/abort the transaction.} with typical transactions' visualization in time-space is shown in Fig.~\ref{fig:sds2}. The figure shows the causal happens-before or {\em partial-ordering} relation, with node-2 as transaction manager ({\em trn-mgr}) at centre, along with two other nodes: the node-1 and node-3 as execution nodes. The arrows indiacte partial-ordering of the events, each one of them has an associated logical clock time-stamp vector, represented by three arguments, viz., $(1,1,0)$, where three arguments are for node-1, trn-mgr, and node-3, repectively. The Node-1 and 3 are replica nodes of transaction mamanger. As per partial-ordering relation, $tx$-$abort(0, 1, 0)$ $\leq$ $N_3$~$abort(1, 3, 1)$ $\leq$ $tx$-$abort(1, 4, 1)$ $\leq$ $tx$-$aborted(1, 4, 2)$. However, we cannot say any thing about the order of $tx$-$aborted$ $(2, 4, 1)$ vs. $tx$-$aborted$ $(1, 4, 2)$, and similarly about the events $tx$-$abort(0, 1, 1)$ vs. $tx$-$commit(1, 1, 0)$. The partial-ordering relation '$\leq$' stands for ``precedes or simultaneous"~\cite{Ivan16}.  

To explain the transactions and precedence relations, we consider two transaction from Fig.~\ref{fig:sds2}, 1. $N_1$ commit $(1, 2, 0)$ $\to$ $N_3$ abort $(1, 3, 1)$, and 2. tx-abort $(0, 1, 1)$ $\to$ $N_3$ abort $(1, 3, 1)$. In the first case, the Node-1 sends message to tx-mgr to commit the transaction. This causes to generate timestamp $(1, 2, 0)$, where local clock of node-2 is shown increased. This is followed by message by Node-2 to abort the transaction which is so far in buffer of Node-2. This causes the timestamp at Node-2 to change to $(1, 3, 1)$. The effect is aborting the transaction by Tx-mgr at time (logical) as $(1, 4, 1)$, which get communicated to Nodes 1 and 3, shown at the bottom of the diagram.   

In a typical scenario, the developers and analysts can analyze the partial ordering of the events, and can establish the likely chain of happen-before relations between all the events. This case of causality can be helpful for studying the concurrent behavior of the distributed system. Having the distributed events' log available, it is possible to query this log database for certain events and the pattern of interactions between the nodes, and then obtained results can be visualized to understand and identify the causes of failures and malfunctions and can relate them as cause-effect relationships.{}

\subsection{Visualizing run time Distributed Systems} 

ShiViz is a visualization tool useful for visualizing distributed log directly in a browser, as there is no need for installation of any software. Its working is based on happen before relation between the events, that follow the partial ordering relations, and displays the events similar to that shown in Fig.~\ref{fig:infr}. This kind of visualization is called {\em time-space} diagram, which can be source of important information for analysis of a distributed system in operations, as well can be a source for debugging a distributed-system based on executions and logs. One of the analysis that can be carried out is relative ordering of distributed events, to debug any current behavior of the system, to query the distributed system for certain events, and to identify the structural similarities and differences between groups of executions. A typical scenario of time-space diagram of events is shown in Fig.~\ref{fig:sds2}, which cn serve the goal by visualizing event ordering and communication between nodes~\cite{Shilin}.
Using the distributed events log it is possible to perform keyword search, as well as structured search, such that matching events can be listed. For example, searching for a keyword "send" will highlight all events in the diagram that have value of the action field as "send", while example of structure based query is, e.g., time.

\section{Conclusion} 

A log is created due {\em printf} statement in the software system, which acts on encountering specific condition for which it is written, thus analysing the logs help us in discovering the functioning of system, including its memory, IO, cache, logins, processes, intrusions, and so on. But, we get only those events' logs which were coded as part of the system, and not all. 

The Unix/Linux logs are exhaustive enough to dig up into majority events' history, however, those which are not coded, there is no log available, and these needs separate coding to identify the event and print it in the relevant logs.

This article has demonstrated how to investigate about certain events using built-in tools of Unix/Linux. However, it requires a lot analysis to obtain needed inference in results. The happen-before relation when represented as a directed graph, can provide probabilistic estimates using Bayesian belief networks~\cite{chowdhary20}.  

The performance analysis of distributed system is challenging due to unavailability of happen before relations between all the events. However, it is possible to resolve some of them by maintaining time-stamp vectors of all the nodes as part of database of every node in a distributed system. Visualization of events can helps developers visualize the events and their order, searching for communication patterns, and to identify potential event causality. This helps developers to reason about the concurrency of events in an execution, in distributed system state, and in distributed failure modes, as well as useful in formulating hypotheses about system behavior and verify them via execution visualizations. Many of such tools have limitations as they provide low-level ordering information, which is limitation for understanding high-level system behavior, and its visualization is usually based on logical time-stamp and not real-time ordering, which is not sufficient for study of certain performance characteristics. More advanced tools can be constructed using Bayesian belief networks, which are based on probabilistic estimates, are scalable and can give analysis results on the face of partial and incomplete information.

\end{document}